\documentclass[preprint2]{aastex}

\def\etal{{\it et al. }} 
\def\Msol{\thinspace\hbox{$\hbox{M}_{\odot}$}}

\shorttitle{} 
\shortauthors{} 

\slugcomment{Accepted for publication in ApJ}
 
\begin{document} 

\title
{On the Metallicity of Star-forming  Dwarf  Galaxies}

\author{Francois Legrand,\altaffilmark{1}  
Guillermo Tenorio-Tagle,\altaffilmark{2} 
Sergey Silich,\altaffilmark{2,3}
Daniel Kunth,\altaffilmark{1} and 
Miguel Cervi\~no \altaffilmark{4}
}
\altaffiltext {1}{Institut d'Astrophysique de Paris; 98bis Boulevard Arago, 
75014 Paris, France}
\altaffiltext{2}{Instituto Nacional de Astrof\'\i{}sica Optica y 
Electr\'onica, AP 51, 72000 Puebla, Mexico}
\altaffiltext{3}{Main Astronomical Observatory National Academy of Sciences 
of Ukraine, 03680 Kiyv-127, Golosiiv, Ukraine}
\altaffiltext{4}{Observatoire Midi Pyr\`en\'ees; 14, avenue Edouard Belin, 
31400 Toulouse, France}

\begin{abstract}  We construct three extreme different scenarios  
of the star formation histories applicable to a sample of dwarf galaxies,  
based either on  their present metallicity or their luminosity. 
The three possible scenarios imply different mechanical energy  input rates  
and these we compare with the theoretical lower limits established for the  
ejection of processed matter out of dwarf galaxies. The comparison strongly  
points at the existence of extended gaseous haloes in these galaxies,  
acting as the barrier that allows galaxies to retain their metals and enhance  
their abundance. At the same time our findings strongly point at a continuous  
star-forming process, rather than to coeval bursts, as the main contributors 
to the overall metallicity in our galaxy sample. 
 
\end{abstract} 
 
\keywords{ISM: bubbles, ISM: abundances, general - starburst galaxies}

\section {Introduction}

Given the extreme violence of stellar ejection processes, either through  
supernova explosions or strong stellar winds, the presence of strong reverse  
shocks assures that upon thermalization the ejected matter would reach  
temperatures ($T \sim 10^6 - 10^8$ K) that would strongly inhibit  
recombination, and thus the detection of the newly processed material at  
optical and radio frequencies. Furthermore, it is now well understood  
that  this hot high pressure gas  fills the interior of 
superbubbles and  drives the outer shock that sweeps and accelerates  
the surrounding ISM into an expanding large-scale supershell.  
The continuous energy input rate from coeval starbursts or continuous star  
forming phases, due to last until the last 8 M$_\odot$ star explodes as  
supernova, reassures that the high temperature of the ejected matter is  
maintained above the recombination limit ($T \sim 10^6$ K) allowing  
superbubbles to reach dimensions in excess of 1 kpc. During this time 
the highly metal enriched superbubble gas does not diffuse 
either in the expanding shell nor in its immediate surroundings.  
It has been convincingly argued in recent years that the metallicity detected  
in star-forming galaxies results from their previous history of star  
formation and has nothing to do with the metals presently ejected by their  
powerful starbursts (see Tenorio-Tagle 1996).  
One in fact would have to wait until the end of the supernova phase ($\geq$ 
40 Myr) for the full ejection of all heavy elements synthesized by the 
massive stars. At this stage up to almost 40$\%$ of the starburst mass would 
have been returned to the ISM and heavy elements, such as oxygen, would have 
reached their expected yield value.

Note that the newly processed heavy elements are generated within a  
region of about 100 pc (the typical size of starburst) while upon cooling  
of the superbubble interior and their fall back towards the galaxy,  
the heavy elements are disseminated over a region of several kpc in size. 
After such a large-scale dispersal,  diffusion   leads to  
a thorough mixing of the heavy elements within the host galaxy ISM,  
enhancing its metallicity.    
 
Indeed the final step towards mixing within the ISM is promoted by diffusion 
since  diffusion is largely enhanced in high temperature gases. Therefore, 
if a new major centre of star formation develops and causes the formation of 
an HII region, the heavy elements will rapidly mix to produce a  uniform 
abundance. In the absence of star formation however, the heavy element 
droplets will also diffuse but over larger time-scales (see Tenorio-Tagle 
1996). All together the process of mixing takes several 4 - 6 $\times$ 10$^8$ 
yr to promote an enhanced and  almost  uniform abundance in a region of
up to several kpc in radius around the original starburst.

On the other hand, the observational evidence of  powerful starbursts in  
dwarf galaxies  has led to the idea that, due to their rather shallow  
gravitational potential, supernova  products and even the whole of the  
interstellar medium, may be easily ejected from the host dwarf systems,  
causing the contamination of the intra-cluster medium (Dekel \& Silk 1986;  
De Young \& Heckman 1994). However, the indisputable presence of metals  
(in whatever abundance) in galaxies implies that the supernova products  
cannot be completely lost in all cases. 

This issue has been recently addressed by Silich \& Tenorio-Tagle 
(2001; hereafter referred to as Paper I) and D'Ercole \& Brighenti (1999) who 
show that within a dark matter (DM) and a low-density extended gaseous halo, 
a galactic wind does not easily develop, making the  removal of the galaxy's 
ISM very inefficient. Here we compare the theoretical estimates with some 
well known starbursts in blue compact dwarf (BCD) galaxies. In particular 
we derive three different possible star formation history scenarios 
that assume either a very young coeval starburst or an extended phase of 
star formation of 40 Myr or 14 Gyr, to account for the inferred star 
formation rate (in the first two cases) and the observed metallicity  
(in the third one) from our sample galaxies. For these galaxies we know: 
their HI mass, metallicity and H$\alpha$ or H$\beta$ luminosities, and  
these allow us to infer the energy input rate  expected in each of our 
galaxies, for  each of the three likely histories of star formation 
(see Section 2). Section 3 is devoted to a  comparison of  the energy input 
rates derived for our galaxies with the recent estimates  of the rates 
required to expel either metals or the ISM from galaxies containing a total 
ISM mass in the range of 10$^6$ to 10$^{10}$ M$_{\odot}$ (see Paper I). 
Section 4 gives a discussion of our results.

\section {The Star forming history of BCD galaxies} 
 
Our sample of dwarf galaxies (see Table I)  was selected from Papaderos et al  
(1996), Mas-Hesse \& Kunth (1999), Marlowe, Meurer \& Heckman (1999),  
Lequeux et al. (1979). We took care to exclude galaxies which could be tidal  
dwarfs, because in such a case their metallicity would not be the result of  
the evolution of the present day system. The sample also includes a few dIrr  
galaxies from Skillman, Kennicutt \& Hodge (1989). In Table 1 the first 
column provides the galaxy identification and columns 2 and 4 give the HI 
mass and  metallicity as reported in the references given in columns 3 and 5,  
respectively. The fact that many galaxies, such as IZw 18 (Vilchez \& 
Iglesias-Paramo, 1998; Legrand \etal 2000; see also van Zee \etal 1998), 
IIZw 40, NGC 4214 (Kobulnicky \& Skillman 1996), NGC 1569 (Devost 
\etal, 1997), and others, present a uniform metal abundance within their HII 
region gas, irrespectively of the chosen slit location, and the arguments  
given in section I regarding the unlikely possibility that the metals come  
from the present generation of stars, have led us to assume that  galaxies  
have been evenly polluted throughout their histories,  up the  levels now  
displayed in their HII region gas.   

We examine three different scenarios of the star formation using 
the observed properties of our sample of dwarf galaxies. 
For our first two models, the observed H$_{\alpha}$ luminosities  
(column 6 of Table 1) were used to estimate the gas  mass converted in stars. 
We derive the $UV$ photon output from the exciting cluster  
using Leitherer \& Heckmann (1995) (hereafter referred to as LH95)  
relation $L(H\alpha)$ erg s$^{-1}$ = 1.36$\times 
10^{-12}N(H^{0})$ s$^{-1}$, and then convert it into the SFR. 
References to the observed H$\alpha$
luminosity are given  in column 7 of Table 1.  These correspond to the total
luminosity obtained from digital photometry on CDD images using an H$\alpha$
filter.  Whenever the H$\alpha$ luminosity of a galaxy was not available,
we used H$\beta$ instead and using the
relation: H$\alpha$/H$\beta$ = 2.857, as in LH95. Galaxies for which
the H$\alpha$ luminosity has been inferred  from their H$\beta$ luminosity 
are marked with an asterisk in the column 7 of Table 1. In these cases, it 
can be assumed that the observed H$\beta$ flux also corresponds to the total 
flux, either because the slit apperture has been chosen large enough for a 
realistic comparison with {\it IUE} data (Mas-Hesse \& Kunth, 1999,
Storchi-Bergmann, Kinney, Challis, 1995), or photometrically derived from 
CDD images (Vilchez, 1995 Gallagher, Hunter \& Bushouse, 1989).

We have first assumed that  
all stars  have resulted from a coeval burst. Under these conditions, it is  
well known that the mechanical input rate through winds and supernovae leads 
to an almost immediate constant value to be maintained throughout the  
supernova phase (4 $\times 10^7$ yr). On the other hand the ionizing flux is  
to remain constant at its maximum value only for the first 3.5 Myr, and then  
decreases as $t^{-5}$ (see Beltrametti \etal 1981). The implication of this  
rapid decay is that after 10$^7$ yr the $UV$ photon output is two orders of  
magnitude smaller than its original value, and thus the HII region phase is  
much shorter than the supernova phase. Here we assume that our clusters are  
indeed coeval and young ($\leq$ 3.5 Myr). The H$\alpha$ luminosity is also in  
this case a direct indicator of the number of ionizing photons. However, to  
infer the mass of the starbursts, the number of ionizing photons should be  
linearly scaled using Figure 37 of LH95, which gives a value of  
$N(H^{0}) \sim 10^{53}$ photons s$^{-1}$ for every $10^{6}$ M$_{\odot}$ of  
stars formed. The corresponding mechanical energy input rate can thus be  
evaluated using the calibration given by these authors for an instantaneous  
burst of 10$^6$ M$_\odot$ (LH95, their Figure 55). The values found for the  
mass in stars and mechanical energy injection rate are given in column 2 and  
3 of Table 2, for each entry. 
 
In this model, our mass estimates will be wrong by a factor of ten if  
the age of our  starbursts would be 6.5 Myr, and by a factor of 100 if they  
all are 10 Myr old. The presence of strong  Wolf-Rayet features in the 
spectra of  many of the sources in the sample, as in IZw 18, 
(see  Kunth \& Ostlin 2000, and references therein)  certifies however 
their young age ($\sim$ 3.5 Myr).  
 
In our second model the present star formation rate was assumed to be  
continuous for the last tens of Myr.  In this case,  
the ionizing photon output reaches a maximum constant value after 3 - 4 Myr  
while the mechanical luminosity of the starburst rises to reach its  
asymptotic maximum value only after 40 Myr (LH95). We thus use their  
continuous star formation model  to derive the SFR from the number of ionizing 
photons ($N(H^0)$) as well as the corresponding energy injection rate, 
assuming a continuous and constant SFR over the last 40 Myr (LH95, Figures 38  
and 56).  The resultant SFR and the corresponding mechanical energy input  
rate for each of our galaxies are given  in columns 4 and 5 of Table 2. 
Note that despite the fact that there is no information on the time since  
star formation began in our galaxies, our assumption that $t \geq $ 40 Myr,  
 leads to  an upper limit for the mechanical energy input rate as 
implied from the derived SFR. 
 
In models 1 and 2, we have assumed that all the ionizing photons produced by  
the stellar clusters are completely used up in the ionization of the HII  
region gas. Note however that if a fraction of these, say 50$\%$,  
manage to freely escape the nebulae, or are absorbed by dust, our final  
estimates are only off by a factor of two. 
 
For our third model we have evaluated the amount of stars required to produce  
the observed oxygen abundance in each galaxy of our sample.  
Because this approach is based on the entire galaxy evolution, we used 
here a closed box model and the instantaneous recycling approximation 
despite of the time delay between the metal ejection and their complete 
mixing with the galaxy ISM. 
In the low metallicity limit (or for large remaining gas fraction 
$\mu$ = $\frac{M_{gas}}{M_{gas}+M_{stars}}$) the ISM gas metallicity Z 
is given by:
\begin{equation} 
Z = - y \, \ln(\mu) 
\end{equation} 
\noindent where $y$ is the net yield (Maeder, 1992). This approximation is 
valid for all the objects in our sample. For oxygen this relation becomes: 
\begin{eqnarray}
      \label{eq:closeboxmodel} 
      & & \hspace{-0.5cm} \nonumber
\frac{O}{H+He+metals} = 
      \\[0.2cm]
      & & \hspace{-0.5cm}
- y_{O} \,  \ln\left(\frac{M_{gas}}{M_{gas}+M_{stars}}\right) 
\end{eqnarray} 
\noindent where $y_{O}$ is the oxygen yield. As we only 
have a measurement of the hydrogen content, we have assumed a constant He/H 
ratio: $M_{gas}=M_{HI+He}=1.34\,M_{HI}$ (Kunth \& Sargent 1986). 
 
Thus, reverting equation \ref{eq:closeboxmodel} we can derive 
the mass of stars required to produce the observed oxygen abundance in our 
objects, as a function of their HI mass and the oxygen  yields. 
 
It is important to notice that the derived mass in stars is a lower limit. 
Indeed, under the assumption of a closed box model, all of the metals 
produced remain within a  galaxy and contribute to globaly enhance its 
metallicity. If some of the metals could escape into the intergalactic space, 
the amount of stars needed to produce the actual measured metallicities will 
have to be larger. Of course, this mass also depends on the yields used, 
and the normalization of the yields depends on the adopted initial mass
function. Here we used the oxygen yield Y$_O$ = 0.0026 computed from 
Maeder (1992) models for an initial mass function with a power law index 
-2.35 between 0.1 $M_{\odot}$ and 120 $M_{\odot}$. The evaluated stellar 
mass in the range 1-120 $M_{\odot}$ needed to produce the observed global 
metallicity of our galaxies is given in column 6 of Table 2.  
 
In the last scenario we have assumed that star formation has taken place all  
over the volume of the host galaxies. We have then  divided the derived  
stellar masses by 14 Gyrs to compute the continuous star formation rate  
required to produce the observed metallicity over a Hubble time. This star  
formation rate is given in the  column  7 of Table 2.  
Note that this scenario   
leads to a lower limit for the mechanical energy injection rate. Indeed, if 
the same amount of stars were formed during burst events, the expected  
mechanical energy input rate should become momentarily larger. 

The corresponding mechanical energy injection rate (column 8 of Table 2) was 
computed using the normalization $log(L_{mech}$) erg s$^{-1}$ = 41.8. 
This is a mean value predicted by a continuous starburst model, for different
star metallicities, after the  equilibrium regime is reached (see Figure 56 
of LH95)
under the assumption of a Salpeter initial mass function between 1 and 
100 M$_{\odot}$ and star formation rate of 1 M$_{\odot}$ yr$^{-1}$.
Therefore we have used this value for our sample regardless of the galaxy 
metallicity.

Note that if a continuous star formation process, spread over the whole  
galaxy volume and over a Hubble time, is required to produce the observed  
global abundances detected in our galaxies, then the implied SFRs and total  
mechanical luminosities are comparable to the values derived for either our  
model 1 or 2. This is remarquable since for models 1 and 2,
the SFR, the  mass in stars, and  the mechanical luminosities 
were inferred independently  from the present--day  galaxy luminosity.  
Although metals presently being produced are  
hidden from view (at least in the optical regime) because  the  
recombination process is inhibited at  high temperatures  
when crossing the reverse shocks (see section I),  
we can derive, for each galaxy, the number of "typical" starburst events 
similar to the present ones, required  to account for the  
observed metallicities. If one divides the stellar mass implied by chemical  
evolution (column 6 in Table 2) over the starburst mass derived in  
scenarios 1 and 2, one would find the number of events needed to explain  
the metal content, evenly spread over the galactic volumes. Column 9  of  
Table 2 provides the number of extended (40Myr) or coeval bursts,  
required to account for the metallicity of the galaxies in our sample.  Note 
that the derived numbers of bursts are also a lower limit estimate as the 
closed box model provides a lower limit on the stellar mass. Nethertheless,
the large number of bursts required in most of the cases, both in models 1 
and 2, simply reflects the fact that it is the whole galactic volume what 
aught to be contaminated. These large numbers seem unrealistic and thus one 
is lead to conclude that the main agent causing the global metallicity 
present in galaxies is a uniform, constant and long-lasting SFR, spread 
(as in model 3) over the galaxy volume. Starbursts may take place throughout 
the history of the galaxy, with a star--forming activity similar to what is 
found in the whole galactic volume, however, their contribution to the total 
metallicity would not be the most significant. 
 
Finally, we compared the values of the derived mechanical energy injection 
rate in the three considered cases, with the predictions of the 
hydrodynamical models of McLow \& Ferrara (1999) and Silich \& Tenorio-Tagle 
(2001). The results are shown in Figure 1.

\section {Comparison with the theoretical estimates}

\subsubsection{The energy requirements}

Figure 1 shows our energy estimates resultant from the numerical integration  
of the hydrodynamic equations for a coeval starburst with a constant energy  
input rate during the first 4 $\times 10^7$ yr of the evolution,  acting on  
an ISM density distribution with a total mass in the  range of  
10$^6$ to 10$^{10}$ M$_{\odot}$. Two extreme density distributions were 
assumed: either rotating flattened disks (lower two lines) or spherical 
galaxies without rotation (upper two lines; see Paper I). Clearly disk-like 
models require less energy to eject their metals into the intergalactic 
medium. This is because the amount of interstellar gas that has to be blown 
out of a disk-like galaxy to open a channel for metal ejection into the
intergalactic medium is much smaller than in the spherically symmetric limit, 
where  all of the metal-enriched ISM has to be accelerated  to reach the
galaxy boundary.

In Figure 1, values below each line imply total retention, while the region  
above each line indicates the expulsion of the hot superbubble interior gas  
(with the new metals) out of disk-like distributions, and of the new metals 
and the whole of the ISM in the spherical cases. Each line that separates the  
two regions  marks the minimum energy input rate needed from a coeval  
starburst to reach the outer boundary of a given galaxy with a supersonic  
speed, regardless of the time that the remnant may require to do so  
(see Paper I).  
 
The energy input rates derived for our sample of galaxies (see Table 2) in  
each of the three star formation  scenarios here considered have been  
incorporated in panels 1a - 1c. Note that the energy input rate derived in  
model 3 from the continuous star formation activity able to account for the  
metallicity of the host galaxies, although originally assumed to take place  
all over the galaxy volume is here compared with the theory estimates derived  
for spatially concentrated starbursts.

\section{Discussion}

Based on sound observed facts such as the HI mass, metallicity and  
either the H$\alpha$ or H$\beta$ luminosity of star-forming dwarf galaxies  
we have constructed three different likely scenarios for their star formation  
histories. The various extreme possibilities result from considering  
either young coeval starbursts or extended phases of star formation lasting  
40 Myr and 14 Gyr respectively.  
 
For the coeval starburst scenario, as well as for the second extended and  
constant star formation rate model, the star formation rate and the total 
mass of the starburst have been inferred from the observed H$\alpha$ and  
H$\beta$ luminosity of each of the galaxies. The third scenario assumes a  
constant star formation rate over a Hubble time (14 Gyr), as low as it is  
required to account for the observed metallicity in each of these galaxies.  
Based on the synthetic properties of coeval and extended starbursts  
we have inferred the energy input rate to be expected in each of the  
cases and these have been compared with the recent theoretical estimates of  
the minimum energy input rate required to vent the processed matter into the  
intergalactic medium (see Paper I).  
 
In all three star formation history scenarios, all galaxies in our sample lie  
above the lower limit first derived by Mac Low \& Ferrara (1999),  
and then confirmed in Paper I, for the ejection of metals out of flattened  
disk-like ISM density distributions when energized by relatively low mass  
starbursts. Under these assumptions, the observed metallicity in these  
systems cannot be accounted for. Our results however, place the bulk of our  
sample, in all three considered star formation history  
scenarios, below the minimum energy input rate range of values derived for  
galaxies that present an extended low density gaseous halo. And thus, as  
shown in Paper I, it seems to be the mass of the extended low density halo,  
the one that acts as the barrier against metal ejection into the IGM. 
 
Note that in the 14 Gyr extended star formation case, all galaxies lie below  
the lower limit imposed by the presence of an extended low density halo  
(see Figure 1c). A few objects namely: NGC 1569, IZw 18, Haro 2, and Mrk 49  
surpase this lower limit in the case of a 40 Myr extended star formation  
history (see Figure 1b), as it is also the case of NGC 1569 in the coeval  
starburst scenario (see Figure 1a). These facts imply that  the limit derived  
by Silich \& Tenorio-Tagle (2001) is indeed a lower limit as it does not  
consider all possible phases of the ISM, nor magnetic fields and it has  
(as other models in the literature) assumed a constant energy input rate. 
All of these are issues that if thoroughly considered could substantially  
raise the energy input rate required for mass ejection out of dwarf galaxies  
hence the metals are indeed trapped in all cases. A second possibility is 
that if the metallicity in these galaxies results from several  extended star  
formation phases, or from several  coeval bursts (see column 9 of Table 2),   
the metals produced by the present phase of star formation will only in a  
few of the cases be able to escape into the IGM. 
 
However, given the large number of star formation events that have been  
derived for many of our galaxies in both models 1 and 2, it is obvious that  
the main contribution to the global metallicity measured in star-forming 
dwarf galaxies is due to a long-lasting and continuous star forming process, 
spread over the whole galaxy volume. A few coeval and massive bursts, similar 
to the ones presently energizing the galaxies, may occur throughout history. 
Taking 100 Myr time interval between the sequential starbursts, one can find, 
that in most of the galaxies from oursample, instantaneous starburst may 
account for 1/10 or even much less of the detected metal abundances.

The main implication of our results is that one has to invoke the existence 
of extended  gaseous haloes in order to account for the metals present in 
galaxies. Predicted haloes, despite acting  
as a barrier to the loss of the new metals, have rather low densities  
($<n_{halo}> \sim 10^{-3}$ cm$^{-3}$) and thus have a long recombination  
time ($t_{rec} = 1/(\alpha n_{halo}$); where $\alpha$ is the recombination  
coefficient) that can easily exceed the lifetime of the  HII region  
($t_{HII}$ = 10$^7$ yr) developed by the starburst. In such a case, the  
haloes may remain undetected at radio and optical frequencies  
(see Tenorio-Tagle et al. 1999), until large volumes are collected into the  
expanding supershells. Note that the continuous $\Omega$ shape that  
supershells present in a number of galaxies, while remaining attached to  
their central starburst, as well as  their small expansion velocity 
(comparable or smaller than the escape velocity of their host galaxy) imply  
that the mechanical  energy of the star cluster is plowing into a continuous  
as yet undetected medium. Furthermore,  in the presence of a halo, it is the  
mass of the halo what sets the limiting energy input rate required for mass  
ejection into the IGM, and not the mass of the disk-like component. This  
argument applies to all galaxies whether spirals, amorphous, irregulars or  
dwarfs.  
 
The authors acknowledge  Miguel Mas Hesse and Michel Fioc for fruitful 
comments and  suggestions. We specially acknowledge the useful remarks of 
the anonymous referee that helped to improve the clarity of the paper. GTT 
thanks CONACYT for the grant 211290-5-28501E which allowed for the completion 
of this study, and MC is supported by an ESA postdoctoral fellowship.

\clearpage 

\onecolumn

\begin{table}
      \label{tab1} 
      \caption {Star-forming dwarf galaxies. Observational data}
      \tiny
      \begin{flushleft}
        \begin{tabular}{|l|c|c|c|c|c|c|}
          \hline\tiny
Name &M(HI)         & ref & 12+Log(O/H) & ref &L(H$_{\alpha}$) & ref \\ 
     &($10^8$ \Msol)&     &             &     &erg s$^{-1}$    &     \\
\hline 
(1)  &(2)           & (3) & (4)         & (5) & (6)            & (7)  \\
\hline
Haro-3          &  5.4    &  1  &  8.37  &  7  &  $7.14 \times 10^{40}$ & 9$^*$    \\
Mkn-49/1318     &  1.9    &  1  &  8.28  & 14  &  $3.71 \times 10^{40}$ & 10$^*$   \\
NGC-2915        &  5.9    & 18  &  8.4   &  7  &  $1.30 \times 10^{39}$ & 8        \\
Haro-2          &  4.6    &  1  &  8.4   &  6  &  $1.20 \times 10^{41}$ & 6$^*$    \\
II-Zw-70        &  3.4    &  1  &  8.07  &  6  &  $2.23 \times 10^{40}$ & 6$^*$    \\
I-Zw-18         &  0.26   &  6  &  7.18  &  6  &  $4.29 \times 10^{39}$ & 6$^*$    \\
Haro-14         &  3.2    &  1  &  8.4   & 13  &  $9.03 \times 10^{39}$ & 9$^*$    \\
IC-10           &  13     &  3  &  8.29  &  3  &  $2.70 \times 10^{40}$ & 11       \\
GR-8            &  0.098  &  4  &  7.68  & 15  &  $1.43 \times 10^{38}$ & 9$^*$    \\
DDO-167         &  0.14   &  2  &  7.66  & 16  &  $7.08 \times 10^{37}$ & 11       \\
DDO-53          &  0.54   &  2  &  7.62  & 16  &  $6.30 \times 10^{38}$ & 11       \\
WLM             &  0.56   &  2  &  7.74  & 16  &  $7.40 \times 10^{38}$ & 11       \\
IC-1613         &  0.49   &  2  &  7.86  & 16  &  $5.25 \times 10^{38}$ & 11       \\
Sex-A           &  0.79   &  2  &  7.49  & 16  &  $1.40 \times 10^{39}$ & 11       \\
Sex-B           &  0.44   &  2  &  7.56  & 16  &  $4.07 \times 10^{38}$ & 11       \\
NGC-6822        &  0.66   &  2  &  8.14  & 16  &  $3.89 \times 10^{39}$ & 11       \\
DDO-47          &  2.51   &  2  &  7.85  & 16  &  $2.34 \times 10^{38}$ & 11       \\
NGC-2366        &  7.94   &  2  &  7.96  & 16  &  $1.58 \times 10^{40}$ & 11       \\
NGC-1569        &  0.59   &  2  &  8.16  & 16  &  $2.19 \times 10^{41}$ & 11       \\
NGC-4214        &  10.96  &  2  &  8.34  & 16  &  $3.71 \times 10^{40}$ & 11       \\
NGC-4449        &  13.49  &  2  &  8.32  & 16  &  $7.08 \times 10^{40}$ & 11       \\
NGC-5253        &  2.04   &  2  &  8.27  & 17  &  $1.97 \times 10^{40}$ & 12$^*$   \\
\hline
\end{tabular}
\end{flushleft}
\scriptsize{References:                                                     
       1)       Thuan \& Martin, 1981       
       2)       Karachentsev et al., 1999      
       3)       Lequeux et al., 1979             
       4)       Huchtmeier et al., 2000 
       5)       Contini, 1996b 
       6)       Mas-Hesse \& Kunth, 1999
       7)       Papaderos et al., 1996
       8)       Marlowe et al., 1995                                   
       9)       Gallagher, Hunter \& Bushouse, 1989  
       10)      Vilchez, 1995                     
       11)      Hunter, Hawley \& Gallagher, 1993
       12)      Storchi-Bergmann, Kinney \& Challis, 1995
       13)      Marlowe, Meurer \& Heckman, 1999 
       14)      Contini, 1996a
       15)      Moles et al., 1990    
       16)      Skillman, Kennicutt \& Hodge, 1989
       17)      Terlevich, Terlevich \& Denicolo, 2001
       18)      Becker et al., 1988  
}
\end{table}

\begin{table}
      \label{tab2} 
      \caption {Star-forming dwarf galaxies. Starburst properties}
      \tiny
      \begin{flushleft}
        \begin{tabular}{|l|c|c|c|c|c|c|c|c|}
          \hline\tiny
Name & M$_{SB}$  & L$_{mech}$ & SFR (1-120) &L$_{mech}$ & Mstar (1-120)   & SFR & L$_{mech}$ & N$_{SB}$ \\ 
     &($10^5$ \Msol)  & (erg s$^{-1}$) & (\Msol yr$^{-1}$)&(erg s$^{-1}$) & ($10^5$ \Msol)   & (\Msol yr$^{-1}$) 
     &(erg s$^{-1}$)  & (exten./coev.) \\ 
\hline
(1)  & (2) & (3)  & (4)  & (5) & (6)  & (7) & (8) & (9)   \\
\hline
     & \multicolumn{2}{|c|}{Coeval starburst} & \multicolumn{2}{|c|}{Extended starburst}  
     & \multicolumn{3}{|c|}{Continuous starformation} &      \\
\hline
Haro-3      & 5.25 & 1.68 $\times 10^{+40}$ & 1.66 $\times 10^{-1}$ & 1.05 $\times 10^{+41}$ 
            & 5546 & 3.96 $\times 10^{-2}$ & 2.50 $\times 10^{+40}$ & 83/1056 \\
Mkn-49/1318 & 2.73 & 8.74 $\times 10^{+39}$ & 8.64  $\times 10^{-2}$ & 5.45 $\times 10^{+40}$  
            & 1411 & 1.01 $\times 10^{-2}$ & 6.36 $\times 10^{+39}$ & 41/517 \\
NGC-2915    & 9.56 $\times 10^{-2}$ & 3.06 $\times 10^{+38}$ & 3.02 $\times 10^{-3}$ & 1.91 $\times 10^{+39}$  
            & 6795 & 4.85 $\times 10^{-2}$ & 3.06 $\times 10^{+40}$ & 5616/71091 \\
Haro-2      & 8.82 & 2.82 $\times 10^{+40}$ & 2.79 $\times 10^{-1}$  & 1.76 $\times 10^{+41}$ 
            & 5298 & 3.78 $\times 10^{-2}$ & 2.39 $\times 10^{+40}$ & 47/600 \\
II-Zw-70    & 1.64 & 5.24 $\times 10^{+39}$ & 5.19 $\times 10^{-2}$  & 3.27 $\times 10^{+40}$ 
            & 1291 & 9.22 $\times 10^{-3}$ & 5.82 $\times 10^{+39}$ & 62/788 \\
I-Zw-18     & 3.15 $\times 10^{-1}$ & 1.01 $\times 10^{+39}$ & 9.97 $\times 10^{-3}$ & 6.29 $\times 10^{+39}$ 
            &  10  & 7.10 $\times 10^{-5}$ & 4.48 $\times 10^{+37}$ & 2/32 \\
Haro-14     & 6.64 $\times 10^{-1}$ & 2.12 $\times 10^{+39}$ & 2.10 $\times 10^{-2}$ & 1.33 $\times 10^{+40}$ 
            & 3686 & 2.63 $\times 10^{-2}$ & 1.66 $\times 10^{+40}$ & 439/5552 \\
IC-10       & 1.99 & 6.35 $\times 10^{+39}$ & 6.28 $\times 10^{-2}$ & 3.96 $\times 10^{+40}$  
            & 9997 & 7.14 $\times 10^{-2}$ & 4.51 $\times 10^{+40}$ & 398/5035 \\
GR-8        & 1.05 $\times 10^{-2}$ & 3.36 $\times 10^{+37}$ & 3.32 $\times 10^{-4}$ & 2.10 $\times 10^{+38}$ 
            &  13  & 9.14 $\times 10^{-5}$ & 5.76 $\times 10^{+37}$ & 96/1218 \\
DDO-167     & 5.21 $\times 10^{-3}$ & 1.67 $\times 10^{+37}$ & 1.65 $\times 10^{-4}$ & 1.04 $\times 10^{+38}$  
            &  17  & 1.24 $\times 10^{-4}$ & 7.82 $\times 10^{+37}$ & 263/3335 \\
DDO-53      & 4.63 $\times 10^{-2}$ & 1.48 $\times 10^{+38}$ & 1.47 $\times 10^{-3}$ & 9.25 $\times 10^{+38}$ 
            &  60  & 4.32 $\times 10^{-4}$ & 2.73 $\times 10^{+38}$ & 103/1306 \\
WLM         & 5.44 $\times 10^{-2}$ & 1.74 $\times 10^{+38}$ & 1.72 $\times 10^{-3}$ & 1.09 $\times 10^{+39}$ 
            &  85  & 6.10 $\times 10^{-4}$ & 3.85 $\times 10^{+38}$ & 124/1569 \\
IC-1613     & 3.86 $\times 10^{-2}$ & 1.24 $\times 10^{+38}$ & 1.22 $\times 10^{-3}$ & 7.71 $\times 10^{+38}$ 
            &  103 & 7.33 $\times 10^{-4}$ & 4.63 $\times 10^{+38}$ & 210/2660 \\
Sex-A       & 1.03 $\times 10^{-1}$ & 3.29 $\times 10^{+38}$ & 3.26 $\times 10^{-3}$ & 2.06 $\times 10^{+39}$ 
            &  64  & 4.57 $\times 10^{-4}$ & 2.88 $\times 10^{+38}$ & 49/621 \\
Sex-B       & 2.99 $\times 10^{-2}$ & 9.58 $\times 10^{+37}$ & 9.47 $\times 10^{-4}$ & 5.98 $\times 10^{+38}$ 
            &  42  & 3.03 $\times 10^{-4}$ & 1.91 $\times 10^{+38}$ & 112/1416 \\
NGC-6822    & 2.86 $\times 10^{-1}$ & 9.15 $\times 10^{+38}$ & 9.05 $\times 10^{-3}$ & 5.71 $\times 10^{+39}$ 
            &  310 & 2.22 $\times 10^{-3}$ & 1.40 $\times 10^{+39}$ & 86/1084 \\
DDO-47      & 1.72 $\times 10^{-2}$ & 5.51 $\times 10^{+37}$ & 5.44 $\times 10^{-4}$ & 3.44 $\times 10^{+38}$ 
            &  512 & 3.66 $\times 10^{-3}$ & 2.31 $\times 10^{+39}$ & 2351/29755 \\
NGC-2366    & 1.16 & 3.72 $\times 10^{+39}$ & 3.68 $\times 10^{-2}$ & 2.32 $\times 10^{+40}$ 
            & 2193 & 1.57 $\times 10^{-2}$ & 9.88 $\times 10^{+39}$ & 149/1887 \\
NGC-1569    & 1.61 $\times 10^{+1}$ & 5.15 $\times 10^{+40}$ & 5.10 $\times 10^{-1}$ & 3.22 $\times 10^{+41}$ 
            &  295 & 2.11 $\times 10^{-3}$ & 1.33 $\times 10^{+39}$ & 1/18 \\
NGC-4214    & 2.73 & 8.74 $\times 10^{+39}$ & 8.64 $\times 10^{-2}$ & 5.45 $\times 10^{+40}$ 
            & 10073& 7.19 $\times 10^{-2}$ & 4.54 $\times 10^{+40}$ & 291/3688 \\
NGC-4449    & 5.21 & 1.67 $\times 10^{+40}$  & 1.65 $\times 10^{-1}$ & 1.04 $\times 10^{+41}$ 
            & 11533& 8.24 $\times 10^{-2}$ & 5.20 $\times 10^{+40}$ & 175/2215 \\
NGC-5253    & 1.45 & 4.64 $\times 10^{+39}$ & 4.59 $\times 10^{-2}$ & 2.89 $\times 10^{+40}$ 
            & 1464 & 1.05 $\times 10^{-2}$ & 6.60 $\times 10^{+39}$ & 80/1010 \\
\hline

\end{tabular}

\end{flushleft}

\end{table}

\clearpage
 
\begin{figure}
\plotone{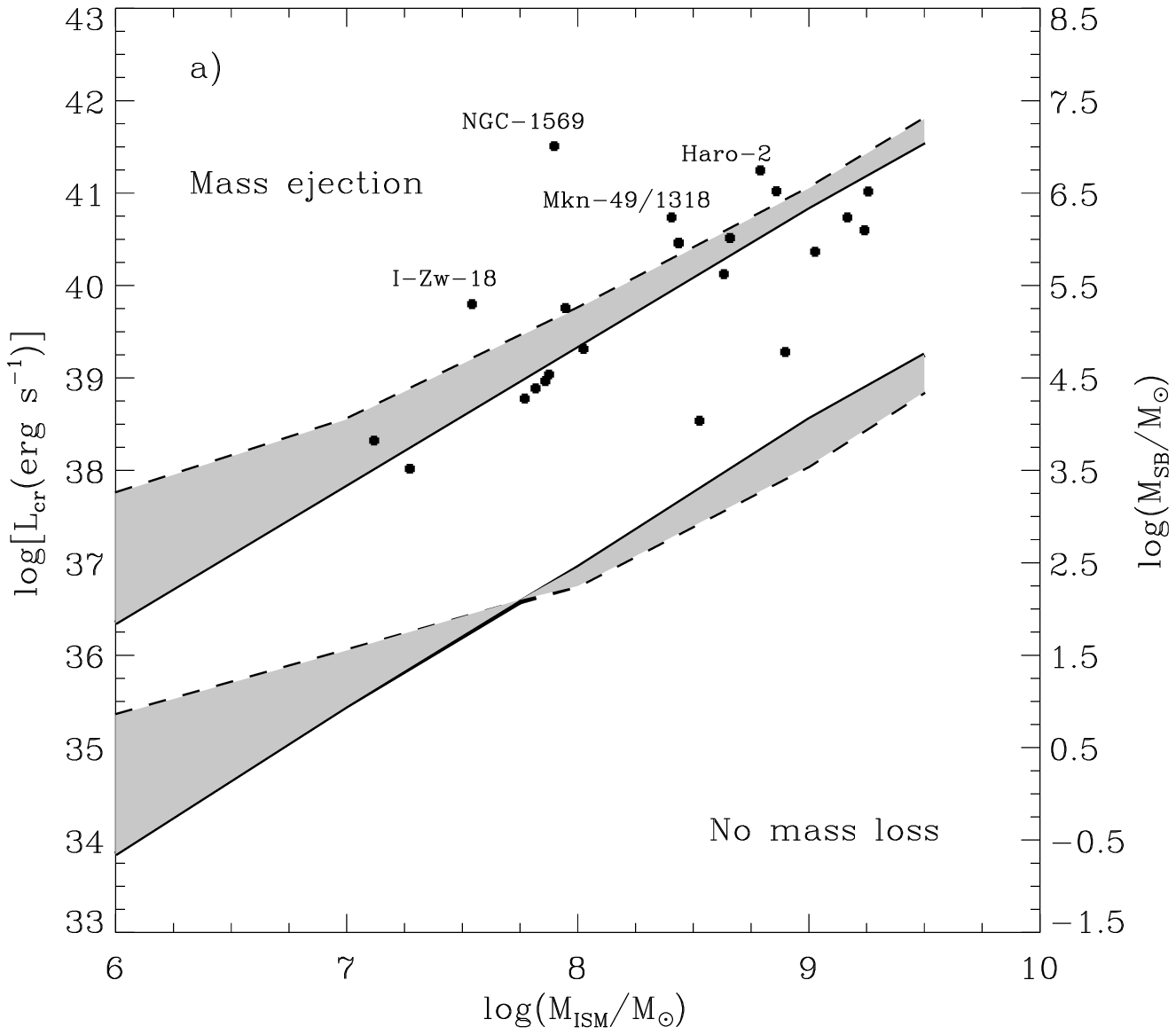} 
\end{figure}

\begin{figure}
\plotone{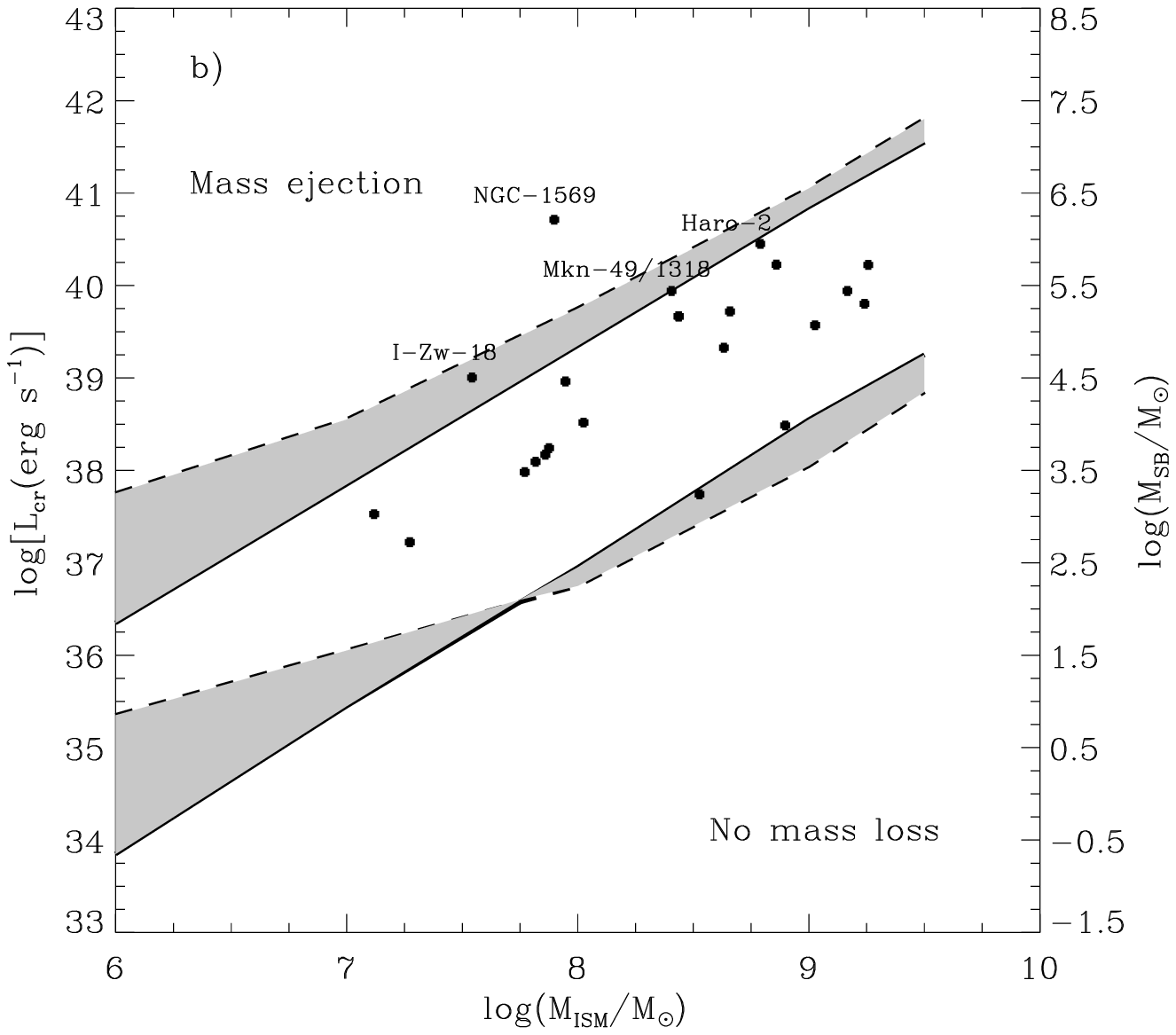} 
\end{figure}

\begin{figure}
\plotone{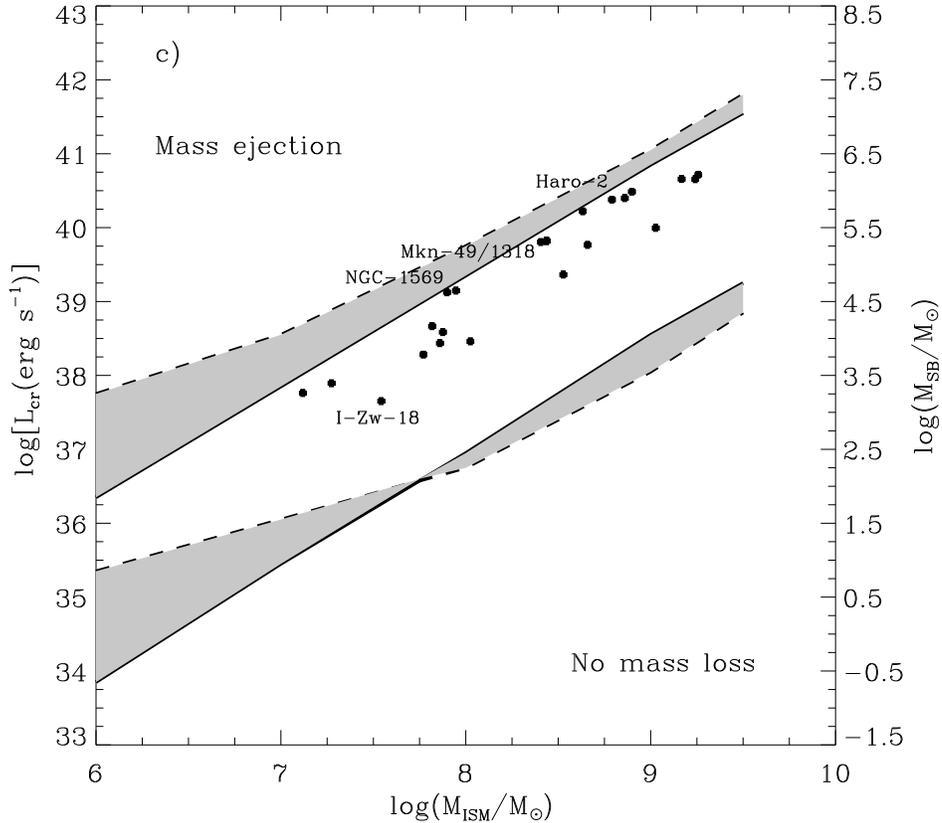}  
\caption
{Energy estimates. The log of the critical mechanical luminosity  
(left-hand side axis) and mass of the star cluster $M_{SB}$  
(right-hand side axis), required to eject matter from galaxies  
with a $M_{ISM}$ in the range 10$^6$ -- 10$^{10}$ M$_\odot$. The lower limit  
estimates are shown for galaxies with extreme ISM density distributions:  
flattened disks (lower two lines), and spherical galaxies without rotation  
(upper lines), for two values of the intergalactic pressure   
$P_{IGM}/k$ = 1 cm$^{-3}$ K (solid lines) and $P_{IGM}/k$ = 100 cm$^{-3}$ K  
(dashed lines). Each line should be considered separately (see Paper I) as  
they divide the parameter space into two distinct regions: a region of no  
mass loss that is found below the line and a region in which blowout and  
mass ejection occur is found above each line. a) -- c) Estimates for 
models 1 -- 3, respectively.} 
\end{figure}

\end{document}